\documentclass[12pt]{amsart}
\usepackage{enumerate}
\usepackage{enumitem}
\usepackage[colorlinks=true, linkcolor=blue, urlcolor=blue, citecolor=blue, anchorcolor=blue, pdfborder={0 0 0}]{hyperref}
\usepackage{breqn}
\usepackage{color,colortbl}
\usepackage{amsthm, amsmath, amssymb}
\usepackage{mathtools}
\usepackage[top=45truemm, bottom=45truemm, left=30truemm, right=30truemm]{geometry}
\usepackage{cancel}
\usepackage{float}
\usepackage{tabularx}
\usepackage{makecell}
\usepackage{array}
\usepackage{ragged2e}
\usepackage{graphicx}
\usepackage[hypcap=false]{caption}
\usepackage{subcaption}
\usepackage{listings}
\usepackage[utf8]{inputenc}
\usepackage{csquotes}
\usepackage{textgreek}
\usepackage[symbol]{footmisc}
\usepackage{tikz}
\usepackage{algorithm}
\usepackage[utf8]{inputenc}
\usepackage[T1]{fontenc}
\usepackage{qcircuit}
\usepackage{algpseudocode}
\usepackage{placeins}
\usepackage{booktabs}

\DeclareCaptionType{referencedlist}[List][List of Referenced Lists]

\definecolor{bg}{rgb}{0.95,0.95,0.95}
\definecolor{LightGray}{RGB}{242,242,242}
\definecolor{LightGreen}{RGB}{208,254,184}

\title[Quantum Cybersecurity]{Enabling Quantum Cybersecurity Analytics in Botnet Detection: Stable Architecture and Speed-up through Tree Algorithms}

\author[Madjid Tehrani]{\href{https://orcid.org/0000-0002-4838-5865}{\includegraphics[scale=0.06]{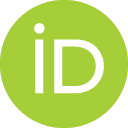}}\hspace{1mm}Madjid Tehrani}
\address{Madjid G. Tehrani\\George Washington University \\Washington, DC 20052, USA}
\curraddr{}
\email{madjid\_tehrani@gwu.edu}

\author[Eldar Sultanow]{\href{https://orcid.org/0000-0001-5257-2236}{\includegraphics[scale=0.06]{orcid.png}}\hspace{1mm}Eldar Sultanow}
\address{Eldar Sultanow\\Capgemini Deutschland GmbH\\Nuremberg, Germany}
\curraddr{}
\email{eldar.sultanow@capgemini.com}

\author[William J Buchanan]{\href{https://orcid.org/0000-0003-0809-3523}{\includegraphics[scale=0.06]{orcid.png}}\hspace{1mm}William J Buchanan}
\address{William J Buchanan\\Edinburgh Napier University\\Edinburgh, UK}
\curraddr{}
\email{b.buchanan@napier.ac.uk}

\author[M.\ Amir]{\href{https://orcid.org/0000-0001-6167-156X}{\includegraphics[scale=0.06]{orcid.png}}\hspace{1mm}Malik Amir}
\address{Malik Amir\\McGill University\\ Université de Montréal\\ Montreal\\Canada}
\curraddr{}
\email{malik.amir.math@gmail.com}

\author[Anja Jeschke]{\href{https://orcid.org/0000-0001-7723-3986}{\includegraphics[scale=0.06]{orcid.png}}\hspace{1mm}Anja Jeschke}
\address{Anja Jeschke\\Capgemini Deutschland GmbH\\Hamburg, Germany}
\curraddr{}
\email{anja.jeschke@capgemini.com}

\author[Raymond Chow]{\href{https://orcid.org/0000-0002-6668-4594}{\includegraphics[scale=0.06]{orcid.png}}\hspace{1mm}Raymond Chow}
\address{Raymond Chow\\George Washington University \\Washington, DC 20052, USA}
\curraddr{}
\email{laserray@gwu.edu}

\author[Mouad Lemoudden]{\href{https://orcid.org/0000-0002-0114-1054}{\includegraphics[scale=0.06]{orcid.png}}\hspace{1mm}Mouad Lemoudden}
\address{Mouad Lemoudden\\Blockpass ID Lab, Edinburgh Napier University\\Edinburgh, UK}
\curraddr{}
\email{m.lemoudden@napier.ac.uk}


\begin{document}

\begingroup
\let\MakeUppercase\relax
\clearpage\maketitle
\thispagestyle{empty}
\endgroup

\begin{abstract}
For the first time, we enable the execution of hybrid machine learning methods on real quantum computers with 100 data samples and real-device-based simulations with 5,000 data samples, thereby outperforming the current state of research of Suryotrisongko and Musashi from 2022 who were dealing with 1,000 data samples and quantum simulators (pure software-based emulators) only. 
Additionally, we beat their reported accuracy of $76.8\%$ by an average accuracy of $91.2\%$, all within a total execution time of 1,687 seconds. 
We achieve this significant progress through two-step strategy: Firstly, we establish a stable quantum architecture that enables us to execute HQML algorithms on real quantum devices. Secondly, we introduce new hybrid quantum binary classification algorithms based on Hoeffding decision tree algorithms. These algorithms speed up the process via batch-wise execution, reducing the number of shots required on real quantum devices compared to conventional loop-based optimizers. Their incremental nature serves the purpose of online large-scale data streaming for DGA botnet detection, and allows us to apply hybrid quantum machine learning to the field of cybersecurity analytics. We conduct our experiments using the Qiskit library with the Aer quantum simulator, and on three different real quantum devices from Azure Quantum: IonQ, Rigetti, and Quantinuum. This is the first time these tools are combined in this manner.
\end{abstract}

\tableofcontents

\section{Introduction} 
\label{sec:introduction}
In the rapidly evolving digital landscape where cyber threats are growing both in sophistication and pervasiveness, maintaining robust cybersecurity measures has taken center stage. While traditional cybersecurity approaches remain effective to a degree, they often struggle to keep up with the constant flood of cyber-attacks \cite{hussain2020review}. In recent years, machine learning has proven to be valuable in various cybersecurity applications. It's been effective in tasks such as intrusion detection, malware classification, and anomaly detection by harnessing automated data analysis and pattern recognition capabilities \cite{martinez2019machine}. Now, the rise of quantum computing is paving the way for even further improvements in cybersecurity analytics.

Quantum computing, renowned for its ability to perform intricate computations at a speed exponentially faster than traditional computers \cite{nielsen2010quantum}, shows promising potential to revolutionize cybersecurity. Quantum machine learning, which has emerged as the intersection of quantum computing and machine learning, leverages the distinctive properties of quantum systems to devise innovative algorithms with the potential to outperform their classical counterparts \cite{biamonte2017quantum}. In this paper, we explore the domain of quantum-enhanced cybersecurity analytics, with a special focus on employing quantum machine learning algorithms for botnet detection - a pressing cybersecurity issue with significant implications for network security \cite{xing2021}. By utilizing the power of quantum computing, we aim to establish a stable architecture and capitalize on the prospective speed enhancement offered by tree algorithms, thereby strengthening the effectiveness and efficiency of botnet detection methods.

The term \textit{Cybersecurity Analytics} \cite{VermaMarchette2020, Mongeau2021} refers to the application of data analysis techniques to cybersecurity.
Much of the literature on this subject takes a practical approach, offering tangible examples and implementable code for cybersecurity solutions \cite{Parisi2019, Das2021, Tsukerman2019}.
However, a term that encapsulates cybersecurity analytics within the context of a quantum system, such as \textit{Quantum Cybersecurity Analytics}, is yet to be fully coined. This is a goal of our present work. In this paper, we introduce \textit{Quantum Cybersecurity Analytics}, or QCA, as a field that employs quantum technology, particularly quantum machine learning, to devise cybersecurity solutions.

We address the challenges and computational demands inherent to quantum machine learning algorithms through the creation of a stable architecture and the adaptation of the Hoeffding tree algorithm for incremental learning \cite{muallem2017hoeffding}.
The current state of the art defined in \cite{Suryotrisongko2022} shows the classification with a hybrid approach of 1000 data samples on a quantum simulator from a botnet dataset with an accuracy of $76.8\%$, whereas the total execution time is not reported. In their study, no signs of any real-device-based simulations or even computations on real quantum devices is shown. We outperform these achievements in the following ways:
\begin{enumerate}
   \item We have extended the maximum sample size from 1,000 to 5,000 data samples in a quantum machine learning method, using real-device-based simulation through the Quantum Hoeffding Tree Classifier (QHTC) algorithm. Our method achieves an average accuracy of 91.2\% and a final-round accuracy of 100\%, all within a total computation time of 1,687 seconds, which is on par with the total execution time observed in locally deployed quantum simulations. 
   \item Furthermore, and for the first time, we implemented various Hybrid Quantum Binary Classification (HQBC) algorithms on actual quantum devices. We managed to process a maximum of 100 randomly fixed data samples, achieving a top accuracy of 59.0\%.
\end{enumerate}
In addition, our work makes the following additional contributions:
\begin{enumerate}
   \item We overcome the pitfalls due to the instabilities of long-running code on three different Azure Quantum Providers by code hardening. 
   \item We apply the batch-wise Hoeffding Tree algorithm instead of the usual loop-wise algorithms relying on gradient descent. 
   \item We compare a diverse set of binary classifiers on real devices, on real-device-based simulations as well as quantum simulators. All experiments are conducted consistently using the IEEE Botnet DGA dataset. 
   \item Quantum Cybersecurity Analytics is made possible. 
\end{enumerate}
The source code implementation is publicly available at \cite{github_source_code_this_paper}. 
The subsequent sections of this paper are organized as follows: 
Section \ref{sec:background} delves into the background of DGA botnets and machine learning for cybersecurity. 
The details of our methodology are described in Section \ref{sec:methodology}. In Section \ref{sec:code_hardening}, the requirements for a stable architecture to run QML algorithms are identified and the quantum-enhanced Hoeffding tree classifier is introduced. In Section \ref{sec:experimental_results}, the experimental results are presented. Finally, Section \ref{sec:conclusion} serves as the conclusion of this paper.

\section{Background}
\label{sec:background}
In this section, we present an overview of the following subjects and delve into the difficulties and possibilities linked to each one: domain generation algorithms (DGAs) botnets, the utilization of machine learning, hybrid quantum machine learning, and quantum machine learning for the detection of botnets through network traffic data.
\subsection{DGA Botnets}
\label{subsec:DGA_Botnets}

A botnet refers to a network of computers infected and controlled by a single attacker, known as the botmaster \cite{Kambourakis2020, Li2008}. Consequently, combatting and addressing botnets has become an important issue, as they have become a prevalent method for launching various internet-based attacks, such as spam, phishing, click fraud, key logging and cracking, copyright infringements, and Denial of Service (DoS) \cite{Lee2008}. The communication topologies that pose the greatest threat to Command-and-Control (C\&C) servers are Domain Generation Algorithm (DGA), peer-to-peer (P2P), and hybrid structures. Our primary focus lies in examining the communication patterns and protocols employed by DGA botnets. Extensive research on these communication patterns has been conducted by the authors of \cite{Vormayr2017}. To detect malicious domains, we adopt domain name detection techniques. The literature offers a comprehensive study of DGA botnets, which serves as a baseline use case for exploring quantum machine learning, given the well-understood patterns of malicious activities in DGA botnets. 

A DGA-based botnet is an advanced form of botnet that exploits a Domain Generation Algorithm (DGA) to generate seemingly random domain names for its command and control (C\&C) infrastructure, such as Mirai \cite{antonakakis2017understanding} and other well-known botnets listed on \href{https://data.netlab.360.com/dga}{Netlab 360}. The primary objective of using a DGA is to create difficulties for security researchers and law enforcement agencies in tracking and dismantling the botnet's C\&C servers, as the generated domain names change regularly. Cybercriminals commonly employ DGA-based botnets to carry out various malicious activities, including spamming, distributing malware, launching DDoS attacks, and stealing data \cite{Vormayr2017}. These botnets have the ability to infect numerous computers and devices, forming a network that serves various illicit purposes. Compared to other types of botnets, DGA-based botnets are known for their resilience and the challenge they pose in terms of tracking and blocking, as they continuously alter their C\&C infrastructure, making it more arduous to disrupt their operations \cite{manasrah2022dga}. The objective of this paper is to evaluate the current capabilities of Noisy Intermediate-Scale Quantum (NISQ) hardware \cite{bharti2022noisy} using a hybrid quantum machine learning approach to detect DGA botnets, and to explore how quantum machine learning (QML) can enhance the functionality of Security Information and Event Management (SIEM), and Security Orchestration, Automation, and Response (SOAR) systems.

\subsection{Machine Learning}
\label{subsec:classical_ml_network_traffic}
Machine learning is a branch of artificial intelligence (AI) that enables software applications to predict outcomes more accurately without the need for explicit programming \cite{martinez2019machine}. Machine learning algorithms use historical data to predict new output values. This allows them to learn from data and improve their performance over time. It is thus a natural choice to consider machine learning to detect botnets by analyzing network traffic data. 

There are various approaches to implementing machine learning, one of which is supervised classification applied to network traffic data. Brezo et al. proposed a supervised classification method for detecting malicious botnet traffic by analyzing network packets \cite{brezo2012}. Piras, Pintor, Demetrio, and Biggio explored techniques for explaining Machine Learning DGA detectors using DNS traffic data and benchmarked different models, including J48 Decision Tree, k-nearest Neighbors, and Random Forest \cite{Piras2022}. Jia, N. Wang, Y.-Y. Wang, and Hu analyzed the traceability and reconstructed the attack path of a botnet control center using an ant colony group-dividing algorithm \cite{Jia2018}. Pérez et al. introduced an approach for proactive detection and mitigation of botnets in 5G Mobile Networks, utilizing software-defined network and network function virtualization techniques \cite{Perez2017}. Onotu, Day, and Rodrigues demonstrated how Neural Nets can recognize shellcode from network traffic data by employing a multi-layer perceptron approach with a back-propagation learning algorithm \cite{onotu2015}. Maniriho, Mahmood, and Chowdhury conducted a survey on malware detection and classification techniques, considering botnets as a subset of malware from a classical computing perspective \cite{Maniriho2022}.

In conclusion, machine learning offers a powerful way to detect botnets. However, challenges that remain are performance degradation over time of ML algorithms as botnets evolve and change their tactics, but also the fact that ML algorithms are susceptible to adversarial attacks. Adversarial attacks are designed to fool ML algorithms into making incorrect predictions. This can be done by injecting malicious traffic into a network that is designed to look like benign traffic. For the purposes of this work, the use of ML for botnet detection gives us a good benchmark to compare and ground our experimentation with quantum machine learning.

\subsection{Hybrid Quantum Machine Learning}
\label{subsec:hybrid_ml_network_traffic}
Hybrid Quantum Machine Learning (HQML) is an innovative approach that combines the power of quantum computing and classical machine learning. HQML algorithms utilize both quantum and classical computers to tackle complex problems that go beyond the capabilities of either technology alone \cite{de2022survey}. Researchers propose two types of hybridization to achieve quantum advantage: vertical hybridization, which involves leveraging quantum devices with a hardware-agnostic low-level design, topology mapping, and error correction routines; and horizontal hybridization, which divides an algorithm into pre-processing, quantum circuit involvement, and post-processing stages to attain quantum advantage through a software-based approach \cite{broughton2020tensorflow}. In the domain of Quantum Cybersecurity Analytics (QCA), Hybrid Quantum Binary Classifiers (HQBC) can be employed to detect various adversarial cyber events, including spam, phishing, spyware, ransomware, and botnets.
In this paper, we demonstrate that Hoeffding Trees outperform classical machine learning binary classifiers and all known quantum approaches in detecting Domain Generation Algorithm (DGA) botnets. Furthermore, we investigate the conditions under which HQBCs can achieve comparable or superior performance compared to classical machine learning models in DGA botnet detection.

As a related work exploring the application of HQML for DGA botnet detection, \cite{Suryotrisongko2022} investigates Hybrid Quantum Deep Learning (HQDL) and Variational Quantum Classifier (VQC) approaches using the IBM quantum infrastructure based on superconducting loops-hardware. The authors simulate the performance of different combinations of key optimizers with variational forms and feature maps. The robustness of HQDL models against adversarial attacks is also examined \cite{suryotrisongko2022adversarial}. The results reveal that a hardened version of the HQDL model can withstand adversarial attacks.
This study on HQML for DGA botnet detection \cite{Suryotrisongko2022} highlights several knowledge gaps that need to be addressed:
\begin{enumerate}
    \item How do other quantum supervised-learning methods perform in detecting DGA botnets?
    \item What is the impact of different qubit approaches, such as trapped ions, silicon quantum dots, topological qubits, and diamond vacancies, on performance and hardware design?
    \item How does the time complexity of different architectural elements influence performance?
\end{enumerate}
This article focuses specifically on the first knowledge gap. In particular, we will evaluate existing HQBCs that play a crucial role in cybersecurity decision systems, including spam detection, anomaly detection, and botnet detection, among others.

\subsection{Quantum Machine Learning}
\label{subsec:quantum_ml_network_traffic}
Researchers have recently investigated a novel approach to intrusion detection by employing quantum machine learning (QML) algorithms \cite{kalinin2022}. The study conducted experiments to demonstrate the effectiveness of QML-based intrusion detection in processing large-scale data inputs with remarkable accuracy (98\%). Notably, the QML approach exhibited twice the speed compared to conventional machine learning algorithms typically used for the same task. These findings highlight the potential of QML approaches to surpass the performance of classical methods in intrusion detection, showcasing their promising capabilities in the field.

\section{Methodology}
\label{sec:methodology}
The methodology section emphasizes the experimental decisions made in this research. The first Subsection \ref{subsec:experiment_design} covers the selection of quantum devices, real-device-based simulators, and quantum simulators utilized for conducting the experiments. The second Subsection \ref{subsec:data_set} provides an explanation for the selection of the IEEE Botnet DGA Dataset, justifying its suitability for the analysis conducted in this research.

\subsection{Selected Platforms}
\label{subsec:experiment_design}
For this research, we opted to use a combination of real quantum devices, real-device-based simulators, and quantum simulators (pure software-based emulators) to reproduce the results reported in the study by Suryotrisongko et al. \cite{Suryotrisongko2020}, which focused exclusively on quantum simulators. Additionally, our experiments were conducted on three Azure Quantum Providers to expand the research scope beyond the utilization of IBM Quantum \cite{Suryotrisongko2020}. The real quantum devices we selected for our experiments were IonQ, Rigetti, and Quantinuum. To perform quantum simulations, we relied on the Qiskit SDK, utilizing Aer for simulations and real-device-based simulations.

The quantum computing configurations used in our experiments are presented in Table \ref{tab:hardware_setup}. The first column introduces a naming convention for referencing the platforms, facilitating better comprehension of the experimental results presented in Section \ref{sec:experimental_results}. Platforms functioning as real quantum devices are denoted by their respective names followed by the letter R. Platforms that combine real quantum devices with simulations, thereby serving as real-device-based simulators, are denoted by their names followed by the letter S.

\begin{table}[htbp]
\centering
\begin{minipage}{\textwidth}
\centering
\begin{tabular}{l|ll}
  Naming & Machine & Device \\ 
  Convention & Name & Mode   \\ 
 \toprule
  Aer             & Qiskit            & Quantum simulator    \\ 
  \hline
  Quantinuum-R    & Quantinuum H1-2           & Real quantum device         \\ 
  \hline 
  Quantinuum-S    & Quantinuum H1-2 Emulator            & Real-device-based simulator      \\ 
  \hline
  Rigetti-R       & Rigetti Aspen-M-3 with Qiskit          & Real quantum device               \\ 
  \hline
  Rigetti-S       & Rigetti QVM           & Real-device-based simulator                \\ 
  \hline
  IonQ-R          & IonQ Aria            & Real quantum device              \\ 
  \hline
  IonQ-R          & IonQ Quantum Simulator & Real-device-based simulator             \\ 
\end{tabular}
\end{minipage}
\caption{Naming conventions for selected platforms shown with their machine name and their device mode (quantum simulator, real-device-based simulator, or real quantum device)} 
\label{tab:hardware_setup}
\end{table}

\subsection{Description of the Dataset}
\label{subsec:data_set}
In this study, we evaluated our findings on DGA botnets using two datasets: the IEEE Botnet DGA Dataset \cite{Suryotrisongko2020, suryotrisongko2021botnet} and the UMUDGA dataset \cite{zago2020umudga}. The UMUDGA dataset consists of 50 malware samples and is suitable for multiple classifications using HQBCs. However, for the purpose of comparing our results to \cite{Suryotrisongko2022}, we focused solely on the IEEE Botnet DGA Dataset in the current experiments. Nonetheless, the UMUDGA dataset may be considered for future investigations.

The IEEE Botnet DGA Dataset comprises a total of 1,803,333 data records. For our experiments, we randomly selected data samples from this dataset. Specifically, we used 1,000 fixed random data samples for quantum simulators, following the approach in \cite{Suryotrisongko2022}, and real-device-based simulators. Additionally, we utilized 100 fixed random data samples for real quantum devices, and a separate set of 5,000 fixed random data samples to test the new algorithm on real-device-based simulators.

As described in \cite{Suryotrisongko2022}, we extracted seven features from the analyzed domain names in the dataset. These features include:

\begin{enumerate}
    \item CharLength: The character length of the domain name.
    \item EntropyValue: The entropy value calculated using Shannon's function with the probability distribution of characters in the domain name.
    \item RelativeEntropy: The distance or similarity of a domain name to the character probability distributions of either Alexa or DGA domain names, measured using the Kullback-Leibler divergence function.
    \item MinREBotnets: The minimum relative entropy with the domain names of DGA botnets.
    \item InformationRadius: The similarity or distance of a domain name to the domains of the ten botnet DGA families, calculated using the Jensen-Shannon divergence function.
    \item TreeNewFeature: A feature generated by a decision tree algorithm that combines the features Entropy, REAlexa, MinREBotnets, and CharLength to train a predictive model.
    \item Reputation: Provides information about the popularity and credibility of the website.
\end{enumerate}

The summarized statistics for these features, including the mean, standard deviation, minimum, median, maximum, skewness, and kurtosis values, are presented in Table \ref{tab:dataset_statistics_1}.

\begin{table}[htbp]
\centering
\begin{tabular}{l|ccccccc}
Feature & Mean & StDev & Min. & Median & Max. & Skewness & Kurtosis \\
\toprule
CharLength & 17.20 & 6.82 & 4.00 & 16.00 & 73.00 & 0.81 & 0.02 \\
EntropyValue & 3.02 & 0.53 & 0.00 & 3.04 & 4.78 & -0.40 & 0.83 \\
RelativeEntropy & 1.66 & 0.82 & 0.20 & 1.55 & 10.10 & 1.63 & 6.91 \\
MinREBotnets & 1.28 & 0.57 & 0.00 & 1.23 & 5.99 & 0.84 & 1.24 \\
InformationRadius & 0.65 & 0.11 & 0

.24 & 0.65 & 1.17 & 0.34 & 0.12 \\
TreeNewFeature & 0.45 & 0.34 & 0.00 & 0.35 & 0.99 & 0.38 & -1.52 \\
Reputation & 81.66 & 54.12 & 0.00 & 64.51 & 436.31 & 0.99 & 0.21 \\
\end{tabular}
\caption{Selected descriptive statistics of the IEEE Botnet DGA Dataset \cite{Suryotrisongko2020} for the seven features according to the Anderson-Darling normality test.}
\label{tab:dataset_statistics_1}
\end{table}

\section{Stable Architecture for Long-running Experiments}
\label{sec:code_hardening}
This section discusses the issues encountered during long-running experiments and presents a stabilized architecture to address these problems. It includes the introduction of a new binary classifier and highlights relevant implementation issues.

\subsection{Reasons for Instability}
\label{subsec:instability_reasons}
The current versions of Qiskit ML classifiers (qiskit-0.41.1 and qiskit-machine-learning-0.5.0), specifically QSVC, Pegasos, VQC, and QNN, have not been tested for compatibility with Azure Quantum Providers such as IonQ, Rigetti, and Quantinuum. Additionally, graceful exception handling has not been implemented. As a result, during the experimentation phase, we frequently experienced instability, including unexpected aborts and missing error messages in long-running notebook sessions. Code hardening revealed the following reasons for instability during experiments on real quantum devices:

\begin{enumerate}
    \item Issues on the real quantum devices
        \begin{enumerate}[label=(\alph*)]
            \item Failure of a single circuit run causing a cascade effect regardless of progress.
            \item Prioritization and scheduling bugs in the task queue.
            \item Maintenance downtime.
            \item Inability to deploy the quantum cloud architecture on a small scale due to insufficient or outdated documentation.
        \end{enumerate}
    \item Issues with the hosted Jupyter notebooks in Azure Quantum workspace
        \begin{enumerate}[label=(\alph*)]
            \item Kernel failure.
            \item Low memory.
            \item Insufficient number of virtual CPUs.
            \item Lack of visibility on progress and log processing.
        \end{enumerate}
    \item Issues in the communication between real quantum devices and notebooks
        \begin{enumerate}[label=(\alph*)]
            \item Authentication and session failures.
        \end{enumerate}
    \item Issues with the Jupyter Notebook on the client side
        \begin{enumerate}[label=(\alph*)]
            \item Termination after a maximum of 24 hours, regardless of CPU or RAM power.
        \end{enumerate}
    \item Issues related to different real devices
        \begin{enumerate}[label=(\alph*)]
            \item Deprecated APIs of Qiskit.
        \end{enumerate}
    \item Issues stemming from the nature of the algorithm
        \begin{enumerate}[label=(\alph*)]
            \item Excessive number of loops.
            \item Lack of code portability.
            \item Inadequate exception handling.
        \end{enumerate}
    \captionof{referencedlist}{Reasons for instability}
    \label{reflist:instability_reasons}
\end{enumerate}

We discovered that the stability of computing and network elements within the architecture is the primary limitation of cloud-based quantum computer delivery. However, none of our experiments on real quantum devices could last longer than three weeks. We were unable to establish a stable TLS connection and authentication for a 1,000 random fixed data sample, leading us to select a reduced sample size of 100 random fixed data points for real quantum devices.

Subsection \ref{subsec:architecture} will present an architecture design that addresses points (1)-(5) in List \ref{reflist:instability_reasons} of instability reasons. Furthermore, Subsection \ref{subsec:speedup_hoeffding} will discuss necessary algorithmic changes to tackle point (6) in List \ref{reflist:instability_reasons}. It is important to note that our experiments running on quantum simulators did not exhibit any instability.

\subsection{Stabilized Architecture}
\label{subsec:architecture}
Our enhanced architecture design addresses the instability reasons (1)-(5) in List \ref{reflist:instability_reasons}. The original architecture that led to instabilities consisted of an Azure real quantum device and an Azure component that involves an Azure Job Management, a storage account and a authentification component. 

\begin{figure}[htbp]
    \centering
    \includegraphics[width=0.9\textwidth]{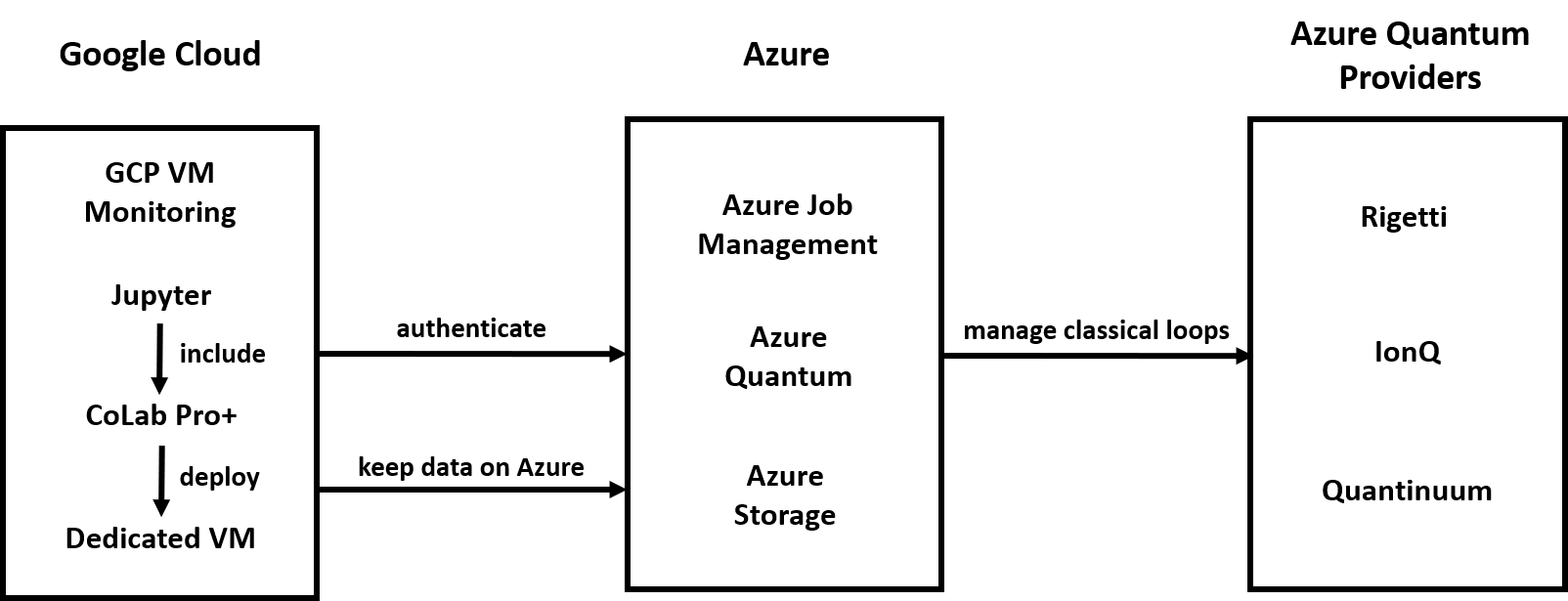}
    \caption{Stabilized architecture of experiments on real quantum devices comprising of three components Google Cloud, Azure and Azure Quantum Providers. }
    \label{fig:architecture_new}
\end{figure}

\begin{figure}[htbp]
\centering
\begin{subfigure}[b]{0.4\textwidth}
\centering
\includegraphics[width=\textwidth]{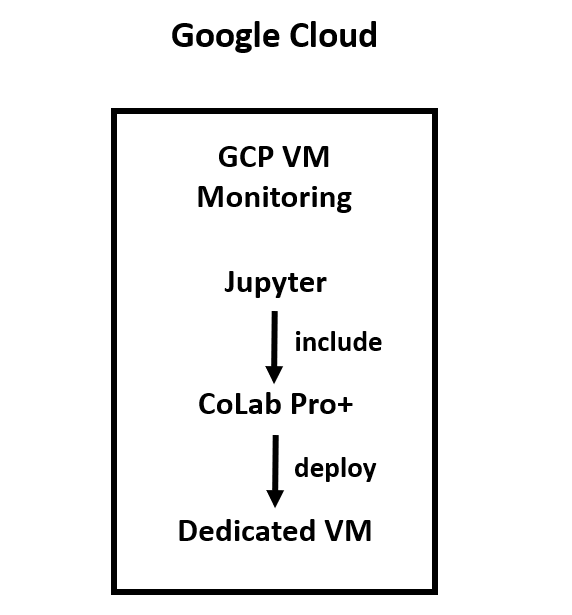}
\caption{(a) Stabilized architecture for \\ QHTC experiment on Aer}
\label{fig:Architecture_stable_v3_QHTC-Aer}
\end{subfigure}
\begin{subfigure}[b]{0.4\textwidth}
\centering
\hfill
\includegraphics[width=\textwidth]{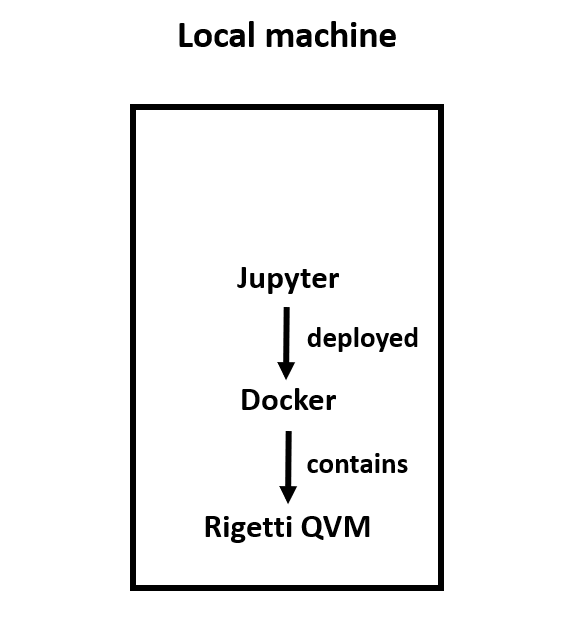}
\caption{(b) Stabilized architecture for \\ QHTC experiments on Rigetti-S}
\label{fig:Architecture_stable_v3_QHTC-Rigetti-S}
\end{subfigure}
\caption{Stabilized architecture for QHTC experiments on quantum simulators Aer and Rigetti-S. The difference in implementation originates from differences in library functionalities available on Aer and Rigetti-S.}
\label{fig:Architecture_stable_QHTC}
\end{figure}

The updated architectures introduce additional components to solve the instability issues mentioned in List \ref{reflist:instability_reasons}. Experiments except QHTC are build on the architecture displayed in figure \ref{fig:architecture_new} and QHTC experiments apply the architecture in figure \ref{fig:Architecture_stable_QHTC}. 

The architecture for experiments except QHTC includes a preceding step in a Google Cloud instance, where a Jupyter and Google Colab Notebook can be deployed on dedicated virtual machines to enable longer runtimes beyond the 24-hour limit. The additional Jupyter Notebook facilitates the implementation of Qiskit code changes for exception handling specific to the algorithm and real quantum device. The Google Colab Pro+ Notebook provides stable runs for more than 1000 random fixed data samples. 
Additionally, a monitoring instance of a GCP virtual machine with diverse logging capabilities aids in identifying, tracking, and resolving errors, including authentication and session failures.

HQML (Hybrid Quantum Machine Learning) opens the door to a new generation of Security Information and Event Management systems known as quantum-enhanced SIEM (QSIEM). To illustrate the functioning of a QSIEM, we present the first use case: defending against Domain Generation Algorithm (DGA) Botnet attacks for DDoS at the application layer using a quantum-enhanced SIEM.
The integration of HQML with a robust SIEM like Azure Sentinel becomes highly beneficial at OSI-layer 7 (application layer), where HTTP and DNS traffic occur. 
This integration enables the detection of malicious domain names generated by DGA-Botnets for command-and-control servers, which are crucial for coordinating DDoS attacks. By identifying and blocking traffic associated with these domains, botnets can be prevented from receiving commands or initiating attack traffic.

\begin{figure}[htbp]
    \centering
    \includegraphics[width=0.5\textwidth]{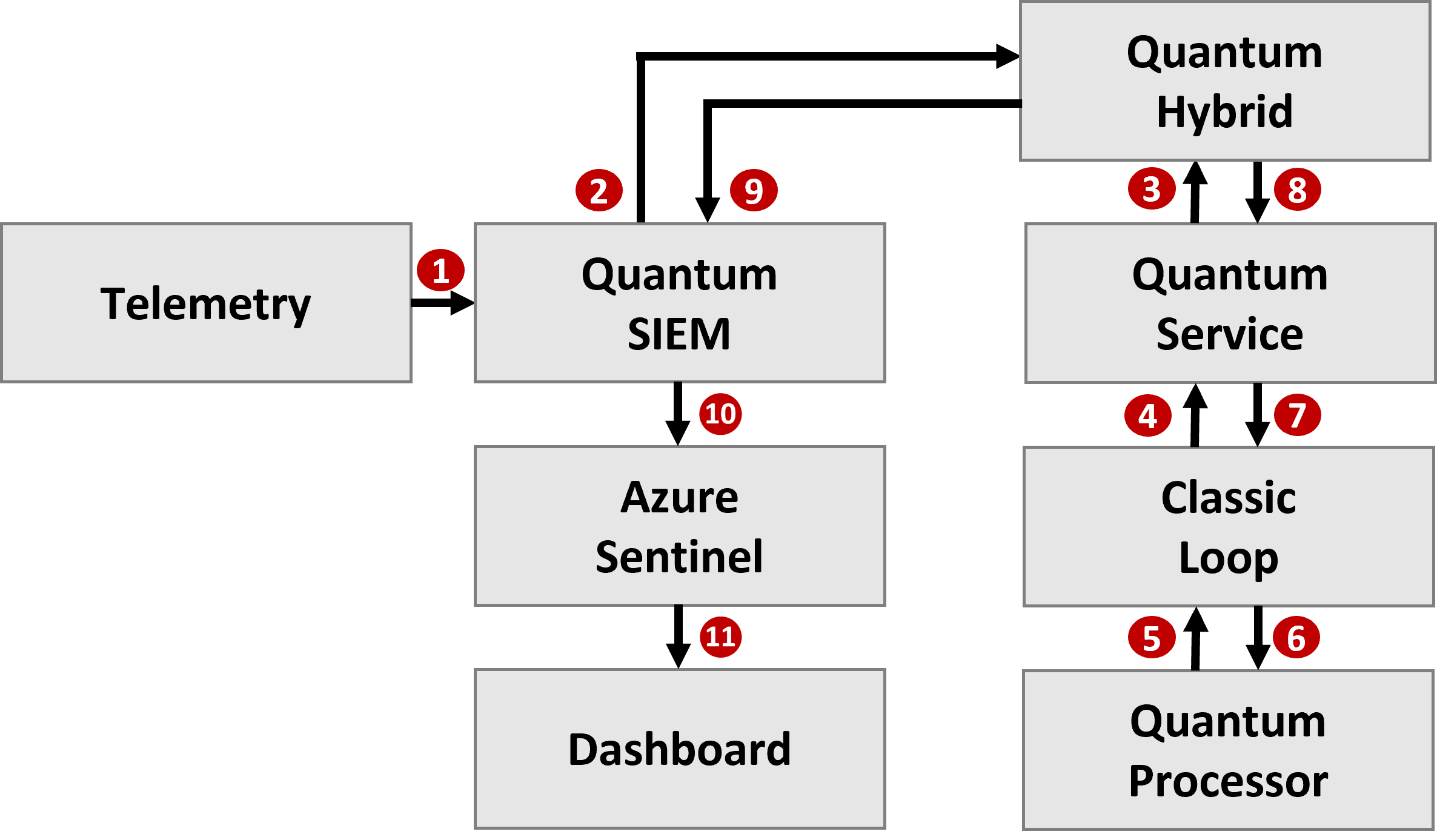}
    \caption{Quantum-enhanced SIEM. The individual steps are marked with numbers in red circles and are explained in List \ref{reflist:architecture_steps}}
    \label{fig:qSIEM}
\end{figure}

Our stabilized architecture aligns with the concept of a quantum-enhanced SIEM solution. The steps in Figure \ref{fig:qSIEM} are explained in List \ref{reflist:architecture_steps}. Steps (2)-(9) are specific to training the HQML algorithm, while the productive algorithm utilizes telemetry input data to generate a classification using Quantum SIEM and Azure Sentinel, which is then displayed on the dashboard.

\begin{enumerate}
    \item Gather and preprocess the telemetry data required for the algorithm described in Subsection \ref{subsec:speedup_hoeffding}.
    \item Perform classic feature engineering as described in Subsection \ref{subsec:data_set}.
    \item Deploy the algorithm for production use on Azure Quantum service.
    \item[(4)-(7)] Execute the entire circuit to and from the real quantum devices using the classical loop.
    \item[(8)] Collect all the results and accumulate the final output.
    \item[(9)] Save and update the classification algorithm.
    \item[(10)] Integrate the classification algorithm with Azure Sentinel.
    \item[(11)] Display the results of the classification algorithm to the user.
    \captionof{referencedlist}{Steps in the solution architecture}
    \label{reflist:architecture_steps}
\end{enumerate}

\subsection{Quantum-enhanced Hoeffding Tree Classifier (QHTC)}
\label{subsec:speedup_hoeffding}
This subsection first describes the historical development of our scientific advances in the direction of the solution, followed by an explanation of the quantum-enhanced Hoeffding Tree Classifier (QHTC).

A realistic QCA solution, i.e., the quantum-enhanced SIEM in \ref{subsec:architecture}, needs to be able to process online big data streaming. Hence, we sought an incremental approach to be applied to already known HQBCs. The most promising algorithmic candidate to reduce execution time and improve accuracy when executed on real-device-based simulators was the PegasosQSVC, in our opinion. Due to its stochastic gradient descent optimizer, the PegasosQSVC performs fewer calculations by iterations and results in better generalization properties of the trained model than conventional gradient descent \cite{SGDvsGD}. Instead of making the PegasosQSVC truly incremental, we applied a batch-wise strategy as an intermediate step between algorithms that need to process the entire training or test data samples at once and incremental algorithms.

\begin{figure}[htbp]
\centering
\begin{subfigure}[b]{0.45\textwidth}
\centering
\includegraphics[width=\textwidth]{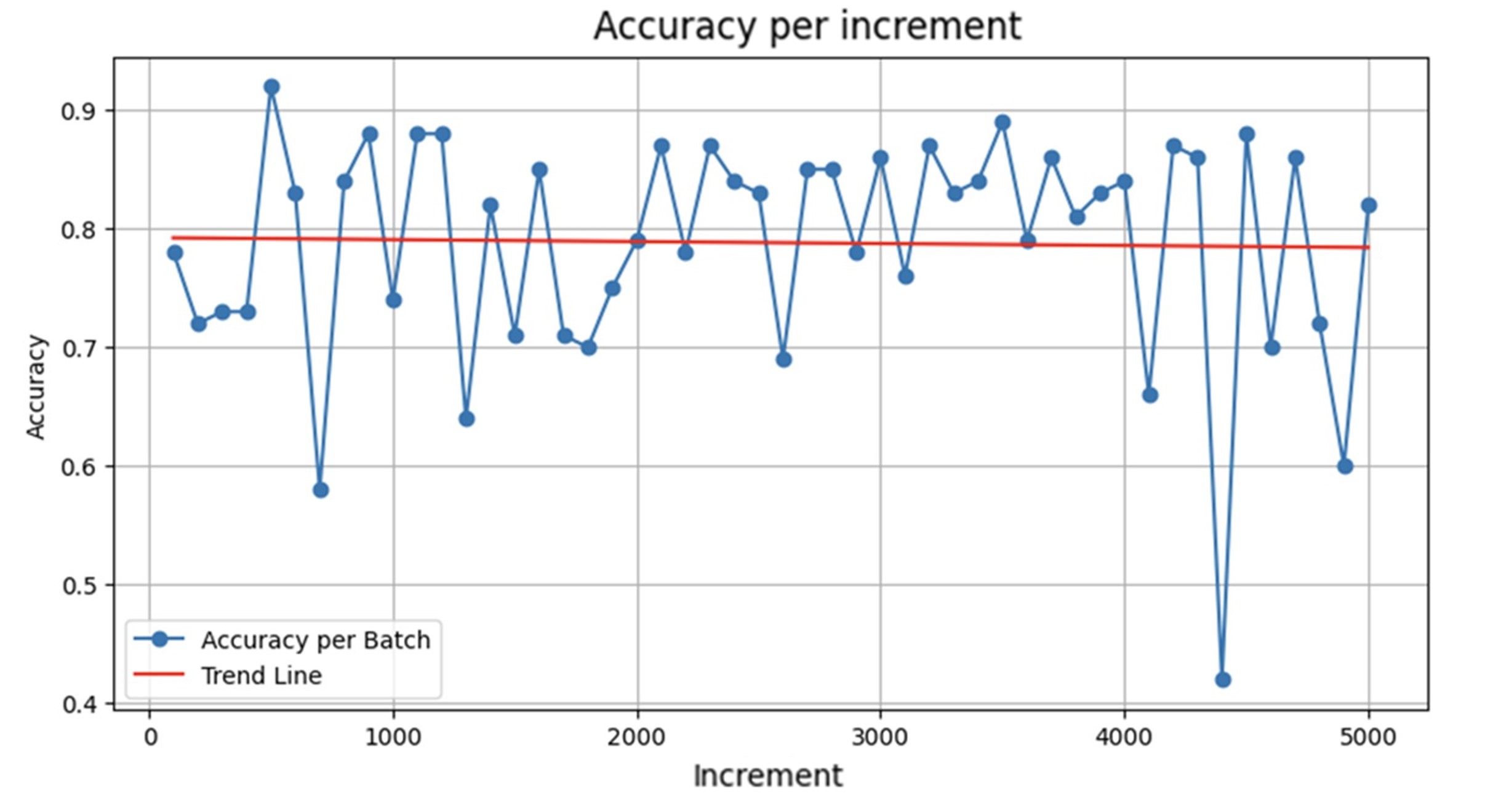}
\caption{(a) Batch size of 100 data samples}
\label{fig:PegasosQSVC_AER_100_accuracy_incremental}
\end{subfigure}
\hfill
\begin{subfigure}[b]{0.45\textwidth}
\centering
\includegraphics[width=\textwidth]{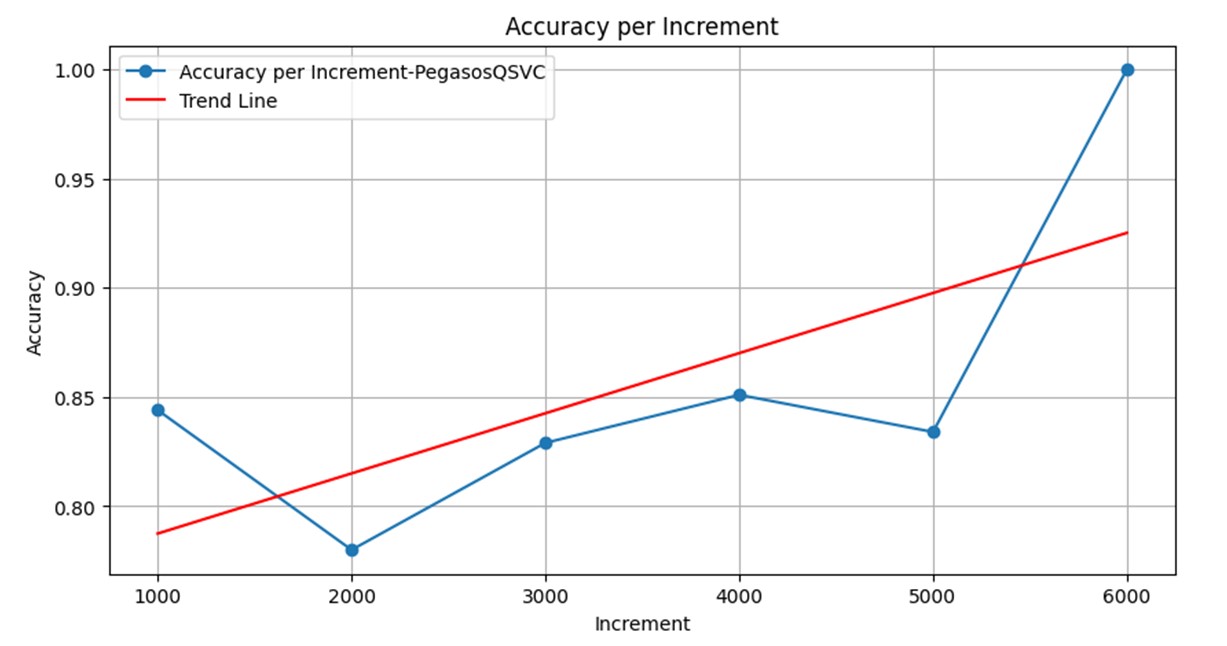}
\caption{(b) Batch size of 1,000 data samples}
\label{fig:PegasosQSVC_AER_1000_accuracy_incremental}
\end{subfigure}
\caption{PegasosQSVC's accuracy on quantum simulator AER with (a) a batch size of 100 data samples does not improve its accuracy with an increased number of batches, unlike (b) a batch size of 1,000 data samples.}
\label{fig:PegasosQSVC_batch_sizes}
\end{figure}

The performance of PegasosQSVC with respect to accuracy development over time is displayed in Figure \ref{fig:PegasosQSVC_batch_sizes} for batch sizes of 1,000 as well as 100 random fixed data samples on the quantum simulator Aer. The PegasosQSVC shows good behavior in terms of accuracy increase with the number of batches if a batch size of 1,000 data samples per batch is applied. But the real quantum devices are not able to handle 1,000 data samples, but only 100 data samples per batch, as the results in Subsection \ref{subsec:results_time_accuracy} will show. In contrast, a batch size of 100 samples will not exhibit the appropriate increase in accuracy on real-device-based simulators or real quantum devices. Smaller batch sizes in the range of 100 data samples require a higher number (one magnitude) of circuits to be sent to the real quantum device, which will extend the execution time to an inappropriate level. This is the dilemma of NISQ-limited data volumes.

Therefore, we decided to transition to a truly incremental algorithm and apply it batch-wise to reduce the number of shots sent to the real quantum device. The accuracy of a truly incremental algorithm will not suffer in this way. This was the breakthrough in terms of the algorithm's accuracy and execution time on real-device-based simulators.

We found the algorithmic solution in a quantum-modified version of an incremental decision tree approach called the Hoeffding tree algorithm \cite{domingos2000mining}, shown in algorithm \ref{alg:HTC}. It is a generation algorithm for incremental decision trees that applies the Hoeffding bound \cite{hoeffding:1963, MarMoo1994hoeffding}. The standard non-incremental version of the decision tree takes all data samples per leaf at once to compute a decision criterion per leaf. In contrast, the incremental version of a decision tree can process one data sample after another. The main advantage of this generation algorithm is that it guarantees, under realistic assumptions, the generation of an asymptotically arbitrarily similar incremental version of a decision tree compared to the same non-incremental version of the decision tree. Simultaneously, it maintains efficient computation speed. Additionally, the Hoeffding bound is independent of the probability distribution of the data samples. However, this implies the disadvantage that the Hoeffding bound, compared to distribution-dependent bounds, requires more data samples to reach the same level of similarity between the incremental version and non-incremental version of the decision tree.

We introduce the abbreviation HTC (Hoeffding Tree Classifier) for the original Hoeffding tree, shown in algorithm \ref{alg:HTC}. Our quantum-modified version is called the quantum-enhanced Hoeffding Tree Classifier (QHTC), as presented in algorithm \ref{alg:incremental_learning_QHTC} and described below. QHTC is a batch-wise learning procedure that applies HTC with modified input data. We apply the HTC in an equivalent version following the HTC implementation of \cite{skmultiflow} that is shown in algorithms \ref{alg:HTC_2} and \ref{alg:HTC_3}. The first step of QHTC is the mapping of the classical features of the input data to the quantum feature space using ZFeatureMap, although other mappings are also possible. Each feature column entry in the feature row represents a data point in quantum space (qubit) on the Bloch sphere and we want to measure the length of the cycle connecting all qubits per feature row. The reason is that the distance between two qubits represents a measure of how distinguishable they are. This cycle length is referred to as a 'quantum walk' in the code. 

The measurement of the cycle length relies on measuring the distance between two qubits on the Bloch sphere. For that, each qubit is converted via wave functions to its density matrix. These density matrices are listed in the same order as the classical feature columns and the trace distance of two density matrices is applied to measure the distance between two qubits that are neighbors on the cycle. 

The cycle length is determined by the order of data points in quantum space and, hence, by the order of the classical features given in the original data set. The determination of a distance metric that allows reordering of feature columns is left for future research. 
The initialization of HTC is performed accordingly. 

The batch-wise computation of an incremental decision tree reduces the number of shots sent to the real quantum device drastically compared to usual loop-based optimizers, while not compromising its accuracy. This provides a solution to the instability reason (6a) mentioned in List \ref{reflist:instability_reasons}. It allows us to deal with the realistic behavior of today's real quantum devices that are prone to instability due to the noise problem inherent in today's NISQ devices. The execution times and the accuracy benefit accordingly, as the results in Section \ref{sec:experimental_results} show in more detail.

\section{Experimental Results}
\label{sec:experimental_results}
The experimental results for different algorithms and quantum devices are presented in the following subsections, focusing on execution time, accuracy, and additional performance metrics for the QHTC algorithm.

\subsection{Execution Time and Accuracy}
\label{subsec:results_time_accuracy}
In this section, we present the experimental results for different binary classifiers, including VQC, PegasosQSVC, QSVC, SamplerQNN, and EstimatorQNN, in terms of accuracy and execution time on quantum simulators, real-device-based simulators, and real quantum devices.

Tables \ref{tab:results_real} and \ref{tab:results_simulated} showcase the accuracy, total computation time ($\text{T}_{\text{total}}$), chosen feature map, and optimizer for various combinations of platforms and algorithms. The experiments on quantum simulators and real-device-based simulators were conducted with at least 1,000 random fixed data samples, while the experiments on real quantum devices used 100 data samples due to computational limitations and instabilities.

\begin{table}[htbp]
\begin{tabular}{lll|lll}
    \small{Algorithm and Platform} & \small{FeatureMap} & \small{Optimizer} & \small{Accuracy [\%]} & \small{$\text{T}_{\text{total}}$ [s]} \\
        \toprule \small{VQC-IonQ-R} & \small{ZFeatureMap} & \small{COBLYA} & 50 & 1,325,133 \\
        \hline \small{PegasosQSVC-IonQ-R} & \small{ZFeatureMap} & \small{SGD} & 41 & 156,156 \\
        \hline \small{QSVC-IonQ-R} & \small{ZFeatureMap} & \small{COBLYA} & 53 & 283,325 \\
        \hline \small{SamplerQNN-IonQ-R} & \small{ZFeatureMap} & \small{COBLYA} & 56 & 956,540 \\
        \hline \small{EstimatorQNN-IonQ-R} & \small{ZFeatureMap} & \small{COBLYA} & 59 & 1,165,819 \\
        \hline \small{VQC-Rigetti-R} & \small{ZFeatureMap} & \small{COBLYA} & 43 & 1,176,879 \\
        \hline \small{PegasosQSVC-Rigetti-R} & \small{ZFeatureMap} & \small{SGD} & 48 & 355,509 \\
        \hline \small{QSVC-Rigetti-R} & \small{ZFeatureMap} & \small{COBLYA} & 39 & 385,153 \\
        \hline \small{SamplerQNN-Rigetti-R} & \small{ZFeatureMap} & \small{COBLYA} & 53 & 1,601,895 \\
        \hline \small{EstimatorQNN-Rigetti-R} & \small{ZFeatureMap} & \small{COBLYA} & 51 & 1,437,085 \\
        \hline \small{VQC-Quantinuum-R} & \small{ZFeatureMap} & \small{COBLYA} & 44 & 972,732 \\
        \hline \small{PegasosQSVC--Quantinuum-R} & \small{ZFeatureMap} & \small{SGD} & 44 & 972,732 \\
        \hline \small{QSVC-Quantinuum-R} & \small{ZFeatureMap} & \small{COBLYA} & 45 & 472,847 \\
        \hline \small{SamplerQNN-Quantinuum-R} & \small{ZFeatureMap} & \small{COBLYA} & 46 & 1,087,789 \\
        \hline \small{EstimatorQNN-Quantinuum-R} & \small{ZFeatureMap} & \small{COBLYA} & 50 & 1,167,143 \\
\end{tabular}
\caption{Performance results in terms of accuracy and total execution time $\text{T}_{\text{total}}$ of real quantum devices, using 100 data samples for all runs. For each algorithm and platform, the choice of the feature map and the optimizer is also shown.}
\label{tab:results_real}
\end{table}

\begin{table}[htbp]
\small
\centering
\begin{minipage}{\textwidth} 
\begin{tabular}{lll|lll}
    \small{Algorithm and Platform} & \small{FeatureMap} & \small{Optimizer} & \small{Accuracy [\%]} & \small{$\text{T}_{\text{total}}$ [s]} \\
    \toprule 
        \small{VQC-Aer} & \small{ZZFeatureMap} & \small{COBLYA} & 54 & 4,240 \\
        \hline \small{PegasosQSVC-Aer} & \small{ZFeatureMap} & \small{SGD} & 90 & 45   \\
        \hline \small{QSVC-Aer} & \small{ZFeatureMap} & \small{COBYLA} & 87 & 3,091  \\
        \hline \small{SamplerQNN-Aer} & \small{ZFeatureMap} & \small{COBYLA} & 76 & 374 \\
        \hline \small{EstimatorQNN-Aer} & \small{ZFeatureMap} & \small{COBYLA} & 84 & 410 \\
        \hline \small{VQC-IonQ-S} & \small{ZFeatureMap} & \small{COBYLA} & 51 & 957,755 \\
        \hline \small{PegasosQSVC-IonQ-S} & \small{ZFeatureMap} & \small{SGD} & 49 & 113,950 \\
        \hline \small{QSVC-IonQ-S} & \small{ZFeatureMap} & \small{COBYLA} & 50 & 178,529 \\
        \hline \small{SamplerQNN-IonQ-S} & \small{ZFeatureMap} & \small{COBYLA} & 59 & 746,992 \\
        \hline \small{EstimatorQNN-IonQ-S} & \small{ZFeatureMap} & \small{COBYLA} & 63 & 780,480 \\
        \hline \small{VQC-Rigetti-S} & \small{ZFeatureMap} & \small{COBYLA} & 46 & 889,708 \\
        \hline \small{PegasosQSVC-Rigetti-S} & \small{ZFeatureMap} & \small{SGD} & 55 & 206,729 \\
        \hline \small{QSVC-Rigetti-S} & \small{ZFeatureMap} & \small{COBYLA} & 45 & 205,877 \\
        \hline \small{SamplerQNN-Rigetti-S} & \small{ZFeatureMap} & \small{COBYLA} & 58 & 656,629 \\
        \hline \small{EstimatorQNN-Rigetti-S} & \small{ZFeatureMap} & \small{COBYLA} & 54 & 955,654 \\
        \hline \small{VQC-Quantinuum-S} & \small{ZFeatureMap} & \small{COBYLA} & 45 & 806,626 \\
        \hline \small{PegasosQSVC-Quantinuum-S} & \small{ZFeatureMap} & \small{SGD} & 49 & 174,416 \\
        \hline \small{QSVC-Quantinuum-S} & \small{ZFeatureMap} & \small{COBYLA} & 49 & 197,871 \\
        \hline \small{SamplerQNN-Quantinuum-S}  & \small{ZFeatureMap} & \small{COBYLA} & 48 & 852,774 \\
        \hline \small{EstimatorQNN-Quantinuum-S} & \small{ZFeatureMap} & \small{COBYLA} & 53 & 716,581 \\
        \hline QHTC-Rigetti-S & \small{ZFeatureMap} & n.a. & 100$^{\footnote{\textit{Already after 3 out of 5 batches}\label{fn_batch}}}$ & 1,687 \\
\end{tabular}
\end{minipage}
\caption{Performance results in terms of accuracy and total execution time $\text{T}_{\text{total}}$ of quantum simulator and real-device-based simulator experiments, using 5,000 data samples for QHTC, and 1,000 data samples for all other algorithms. For each algorithm and platform, the choice of the feature map and the optimizer is also shown. }
\label{tab:results_simulated}
\end{table}

On real quantum devices, the PegasosQSVC performs well in terms of execution time due to its SGD optimizer which tends to converge a little faster than non stochastic optimizers. The PegasosQSVC stands out as the superior binary classifier. As the APIs of feature maps of Qiskit (see for example \href{https://qiskit.org/documentation/stubs/qiskit.circuit.library.ZFeatureMap.html}{ZFeatureMap}) have no endpoint to change the quantum real device, specific implementations are needed for each algorithm. Hence, we didn't intend to compare QHTC over different quantum real devices. We left the implementation of additional coding routines in order to enforce specific real quantum devices and real-device-based simulators for future investigations. 

The PegasosQSVC shows good accuracy (90\%) and very good execution time (45 seconds) on real-device-based simulators. However, the QHTC algorithm outperforms all other binary classifiers in terms of accuracy, achieving perfect accuracy of 100\% already after 3 out of 5 batches. The accuracy is discussed in more detail in subsection \ref{subsec:results_tree}. Furthermore, QHTC exhibits significantly reduced total execution time compared to other algorithms on real-devise-based simulators.  

The experiments conducted on real-device-based simulators and real quantum devices considered a first step, and further improvements and specific implementations for each algorithm on different devices can be explored in future research. Overall, these results demonstrate that it is possible to construct superior algorithms for cloud-based NISQ deployments on real-device-based simulator Rigetti, achieving comparable execution times to quantum simulators while exceeding in terms of accuracy.

\subsection{Performance Metrics of QHTC}
\label{subsec:results_tree}
We show the results of our QHTC (see algorithm \ref{alg:incremental_learning_QHTC}) which is configured to run with five batches containing 1,000 random fixed data samples each. We apply the feature map ZFeatureMap provided by Qiskit. 
Table \ref{tab:qhtc-performance-metrics} demonstrates achievements in terms of accuracy improvement. The increase in accuracy with the number of batches meets our expectations. QHTC yields the same results for all three feature maps. We obtained an average accuracy of $91.2\%$ and a final-round accuracy of $100\%$ for QHTC already after 3 out of 5 batches. We used the same features and the same dataset as \cite{Suryotrisongko2022} to be able to compare our results with theirs. These features are the same features that are available in the entire dataset itself. This may be the reason for such high accuracy. In future research, we can further improve the metric computation to avoid over-fitting and to make it more realistic by applying a PCA analysis as well as using a k-fold cross-validation per batch, with $k=10$ for example.
In addition, the features EntropyValue and RelativeEntropy possess strong predictor properties for the entire dataset. Hence, the same issue will probably not happen to other datasets that don't possess very strong predictor features.

\begin{table}[htbp]
\centering
\begin{tabular}{c|cccc}
Batch & Accuracy [$\%$] & F1-score [$\%$] & AUC [$\%$] \\
\toprule 
1 & $57.1$ & $4.5$ & $51.1$ \\
2 & $99.0$ & $98.8$ & $98.9$ \\
3 & $100.0$ & $100.0$ & $100.0$ \\
4 & $100.0$ & $100.0$ & $100.0$ \\
5 & $100.0$ & $100.0$ & $100.0$ \\
Average & $91.2$ & $80.7$ & $90.0$ \\
\end{tabular}
\caption{Metric results in terms of accuracy, F1-score and AUC for algorithm QHTC, displayed for 5 batches with 1,000 data samples each and their average. }
\label{tab:qhtc-performance-metrics}
\end{table}

\section{Conclusion and Future Work}
\label{sec:conclusion}
Cybersecurity Analytics involves the collection of data to gather evidence, construct timelines, and analyze threats, thereby enabling the design and execution of a proactive cybersecurity strategy that detects, analyzes, and mitigates cyber threats. The next-generation Quantum Cybersecurity Analytics utilizes hybrid quantum machine learning (HQML) to monitor network activity, promptly identify resource use or network traffic changes, and address threats. This advancement paves the way for a new generation of Security Information and Event Management systems called quantum-enhanced SIEM (QSIEM). 
To illustrate how a QSIEM operates, we presented the first use case of defending against DGA botnet attacks for DDoS at the application layer using a quantum-enhanced SIEM.

As cybersecurity is built upon the analysis of amounts of big data, today's NISQ era poses an obstacle for quantum-enhanced SIEM for cybersecurity due to its inherent instabilities that enlarge with repeated and prolonged computations. This study found a way to overcome parts of the problem by proposing a new form of HQML binary classifiers that lead to significant improvements in the result's accuracy as well as the algorithm's execution times with real-device-based simulations compared to previous algorithms. The breakthrough was the application of a quantum-enhanced version of the incremental Hoeffding tree algorithm in a batch-wise version in order to take account of large amounts of incoming online stream data in addition to responding to the need for a reduced number of shots to the real quantum device. 
In addition to the improved accuracy, the experimental run times in real-device-based simulations were reduced drastically by three orders of magnitude to be in the same order as with the previous algorithms on the quantum simulator Aer that is deployed locally. 

In general, the world of quantum simulators is much more beautiful than the world of computations on real quantum devices. 
This study showed for the first time that HQML algorithms were able to run stably with 100 random fixed data samples for several weeks on Azure Quantum Providers Rigetti, Quantinuum, and IonQ together with the library Qiskit. It is the first time these tools were combined. We achieved this by code hardening throughout the entire data flow process from the Jupyter Notebook to the real quantum devices, including all communications and algorithm-specific implementations of APIs per real quantum device. However, future research needs to build upon our progress in order to make the quantum computations on real devices stable for a much larger portion than 100 random fixed data samples, being just a very small fraction of the entire IEEE Botnet DGA Dataset. The enlargement of stability may also be pursued in the case of quantum simulations, as we only used a random fixed sample size of 1,000 in the usual HQBC case and a random fixed sample size in the QHTC case when conducting real-device-based simulations. 

Moreover, we left the implementation of additional coding routines in order to enforce all specific real quantum devices or real-device-based simulators in the case of the quantum-enhanced version as well as the original version of the Hoeffding tree algorithm for future investigations. In addition, the determination of a distance metric for QHTC that allows reordering of feature columns is left for future research. 
Our focus of this study in this regard was to show the excellent properties of these HQBC algorithms for the DGA botnet classification problem in which we succeeded. 

For future research, we also suggest investing more into PegasosQSVC because if we combine quantum supervised learning with rewarding and quantum reinforcement learning, we may have groundbreaking cybersecurity tools. Because current NISQ and hybrid models can support up to 5,600 qubits, perhaps we don't have a 5,600 network feature in cyber data. Resulting from that, even in this NISQ period, we can probably make strong cyber use cases for existing quantum computers and HQML. 

Furthermore, it is an open question as to what practical problem of which scientific fields the same approach of quantum-enhanced Hoeffding tree algorithms might apply as well. The UMUDGA dataset may be a next suitable choice for the DGA botnet detection field. We elaborated on a number of features of the IEEE Botnet DGA Dataset in order to give researchers from other fields a good starting point for their investigations. 
\\
\\
\textbf{Acknowledgements.} We acknowledge support from Microsoft’s Azure Quantum for providing credits and access to the IonQ, Quantinuum and Rigetti systems used in this paper.

\newpage
\appendix

\newpage
\section{Pseudo-code Algorithms}
\label{appx:algorithms}

\begin{algorithm}[H]
  \caption{\footnotesize The Original Hoeffding Tree Algorithm \cite{domingos2000mining} (HTC)}
  \label{alg:HTC}
  \scriptsize 
  \textbf{Inputs:}
  \begin{itemize}
    \item $S$ is a sequence of examples.
    \item $\mathbf{X}$ is a set of discrete attributes.
    \item $G(\cdot)$ is a split evaluation function.
    \item $\delta$ is one minus the desired probability of choosing the correct attribute at any given node.
  \end{itemize}
  
  \textbf{Output:}
  \begin{itemize}
    \item $HT$ is a decision tree.
  \end{itemize}
  
  \begin{algorithmic}[1]
    \Procedure{HoeffdingTree}{$S, \mathbf{X}, G, \delta$}
      \State Let $HT$ be a tree with a single leaf $l_1$ (the root).
      \State Let $\mathbf{X}_1 = \mathbf{X} \cup \{X_{\emptyset}\}$.
      \State Let $\bar{G}_1(X_{\emptyset})$ be the $\bar{G}$ obtained by predicting the most frequent class in $S$.
      
      \For{each class $y_k$}
        \For{each value $x_{ij}$ of each attribute $X_i \in \mathbf{X}$}
          \State Let $n_{ijk}(l_1) = 0$.
        \EndFor
      \EndFor
      
      \For{each example $(\mathbf{x}, y_k)$ in $S$}
        \State Sort $(\mathbf{x}, y)$ into a leaf $l$ using $HT$.
        \For{each $x_{ij}$ in $\mathbf{x}$ such that $X_i \in \mathbf{X}_l$}
          \State Increment $n_{ijk}(l)$.
        \EndFor
        \State Label $l$ with the majority class among the examples seen so far at $l$.
        
        \If{the examples seen so far at $l$ are not all of the same class}
          \For{each attribute $X_i \in \mathbf{X}_l - \{X_{\emptyset}\}$}
            \State Compute $\bar{G}_l(X_i)$ using the counts $n_{ijk}(l)$.
          \EndFor
          
          \State Let $X_a$ be the attribute with the highest $\bar{G}_l$.
          \State Let $X_b$ be the attribute with the second-highest $\bar{G}_l$.
          \State Compute $\epsilon = \sqrt{\frac{R^2}{2} \cdot \ln\left(\frac{1}{\delta}\right) \cdot \frac{1}{\sum_{ijk} n_{ijk}(l)}}$.
          
          \If{$\bar{G}_l(X_a) - \bar{G}_l(X_b) > \epsilon$ and $X_a \neq X_{\emptyset}$}
            \State Replace $l$ by an internal node that splits on $X_a$.
            \For{each branch of the split}
              \State Add a new leaf $l_m$, and let $\mathbf{X}_m = \mathbf{X} - \{X_a\}$.
              \State Let $\bar{G}_m(X_{\emptyset})$ be the $\bar{G}$ obtained by predicting the most frequent class at $l_m$.
              \For{each class $y_k$ and each value $x_{ij}$ of each attribute $X_i \in \mathbf{X}_m - \{X_{\emptyset}\}$}
                \State Let $n_{ijk}(l_m) = 0$.
              \EndFor
            \EndFor
          \EndIf
        \EndIf
      \EndFor
      \State \textbf{return} $HT$.
    \EndProcedure
  \end{algorithmic}
\end{algorithm}

\begin{algorithm}[H]
  \caption{\footnotesize The HoeffdingTreeClassifier (HTC) following implementation \cite{skmultiflow}}
  \label{alg:HTC_2}
  \scriptsize 

    \begin{algorithmic}[1]
    \Procedure{init}{$nFeatures$, $nClasses$, $delta=0.01$, $tiethreshold=0.05$}
          \State Store the input variables. 
          \State $root$ $\gets \Call{TreeNode}{delta}$
    \EndProcedure

    \vspace{0.25cm}
    \Function{predict}{$X$}
        \State Predict the class labels for the input instances $X$.
        \State \Return $yPredict$
    \EndFunction

    \vspace{0.25cm}
    \Procedure{partialFit}{$X, y$}
        \State Update the tree with new training instances $X$ and their corresponding class labels $y$.
    \EndProcedure

    \vspace{0.25cm}
    \Procedure{updateStatistics}{$X, label$}
        \State Update the statistics of the tree nodes based on the input instance $X$ and its class label $label$.
    \EndProcedure

    \vspace{0.25cm}
    \Procedure{attemptSplit}{$node$}
        \State Attempt to split the given node $node$ based on the Hoeffding bound gain.
    \EndProcedure

    \vspace{0.25cm}
    \Procedure{splitNode}{$node$}
        \State Split the given node $node$ by selecting the best attribute based on the Hoeffding bound gain.
    \EndProcedure
    \vspace{0.25cm}
    \Function{hoeffdingBound}{$node$}
        \State Split the given node $node$ by selecting the best attribute based on the Hoeffding bound gain.
        \State \Return $epsilon$
    \EndFunction
    
    \end{algorithmic}
\end{algorithm}

\begin{algorithm}[H]
  \caption{\footnotesize TreeNode (as part of algorithm \ref{alg:HTC_2})}
  \label{alg:HTC_3}
  \scriptsize 

    \begin{algorithmic}[1]
    \Procedure{init}{$delta$}
          \State Store the input variable.
          \State Initialize further variables. 
    \EndProcedure

    \vspace{0.25cm}
    \Function{isLeaf}{}
          \State Check if the node is a leaf (no children). 
          \State \Return not $children$
    \EndFunction

    \vspace{0.25cm}
    \Procedure{computeErrorRate}{}
          \State Compute the error rate of the node based on the class distribution.
    \EndProcedure

    \vspace{0.25cm}
    \Procedure{computeBestSplittingAttribute}{$nClasses$, $nFeatures$}
          \State Compute the best attribute to split on based on the Hoeffding bound gain.
    \EndProcedure

    \vspace{0.25cm}
    \Function{computeHoeffdingBoundGain}{$classCountsPerValues$, $totalSamplesPerValue$, $nClasses$, $totalSamples$}
          \State Compute the Hoeffding bound gain for the given attribute values and their class counts. 
          \State \Return $gain$
    \EndFunction

    \vspace{0.25cm}
    \Function{hoeffdingBound}{$errorRate$}
          \State Compute the Hoeffding bound for the given error rate $errorRate$. 
          \State \Return $hoeffdingBound$
    \EndFunction

    \vspace{0.25cm}
    \Function{getMajorityClass}{}
          \State Get the majority class label based on the class distribution. 
          \State \Return $majorityClass$
    \EndFunction

    \end{algorithmic}
\end{algorithm}

\begin{algorithm}[H]
  \caption{\footnotesize Batch-wise Learning with the Quantum-enhanced Hoeffding Tree Classifier (QHTC)}
  \label{alg:incremental_learning_QHTC}
  \scriptsize 

    \begin{algorithmic}[1]
    \Procedure{RunQHTC}{}
          \State Load the data set.  
          \State Extract the features and labels from the data set.
          \State Normalize the features using a standard scaler. 
          \State Initialize a few variables. 
          
          \For{each feature row $r$ in data set}
          \State Compute quantum walk distance \Call{QHTCDistanceRigetti}{$r$}. 
          \EndFor
          
          \State Store the new quantum data rows as a combination of the quantum walk distances and labels.
    
          \State Initialize number of features $nFeatures$ with 1. 
          \State Initialize number of classes $nClasses$ with number of unique values in labels.
          \State Initialize function \Call{HoeffdingTree.init}{$nFeatures$, $nClasses$}. 
          
          \State Set the percentage for quantum test data. 
          \State Set the batch size. 
          
          \For{all feature rows with step batch size}
          \State Split the quantum data into training and testing data.
          \State Fit the function \Call{HoeffdingTree.partialFit}{$XTrain, yTrain$} to the quantum training data. 
          \State Predict labels for the quantum test data using the function \Call{HoeffdingTree.predict}{$XTest$}. 
          \State Calculate all performance metrics of the predicted labels $yPredict$. 
          \EndFor
          \State Plot performance metrics per batch. 
    \EndProcedure
    \end{algorithmic}
    
    \vspace{0.25cm}
    \begin{algorithmic}[1]
    \Function{QHTCDistanceRigetti}{$r$}
        \State Connect to Rigetti-S using library pyquil or to Aer using library qiskit. 
        \State Create a wave function simulator $ws$.
        \For{each data point $p$ in feature row $r$}
        \State Compute qubit $q$ as ZFeatureMap transformation of $p$ using library pyquil (Rigetti-S) or qiskit (Aer).
        \EndFor
        \For{each qubit $q$}
        \State Get the wave functions $f(q)$ from $ws$.
        \EndFor
        \For{each wave function $f(q)$}
        \State Compute the density matrices $m_{f(q)}$.
        \EndFor
        \For{$i \gets 1$ to $n$ while $i<n$ and n is length of feature row $r$}
        \State Let single distances $d_i = \Call{TraceDistance}{m_{f(q)}(i), m_{f(q)}(i+1)}$.
        \EndFor
        \State Let single distance $d_n = \Call{TraceDistance}{m_{f(q)}(n), m_{f(q)}(1)}$.
        \State \Return Sum of single distances
    \EndFunction
    \end{algorithmic}

    
    \vspace{0.25cm}
    \begin{algorithmic}[1]
    \Function{TraceDistance}{$m_1$, $m_2$}
        \State Let $d = m_1 - m_2$.
        \State Compute singular values $s_i$ of $d$. 
        \State \Return $0.5*\sum_i |s_i|$
    \EndFunction
    \end{algorithmic}
    
\end{algorithm}

\newpage

\section{Glossary}
\label{appx:glossary}

\begin{table}[htbp]
\centering
\setlength{\tabcolsep}{1.2em}\setlength\extrarowheight{3pt}
\begin{tabular}{l|l}
\textbf{Term} &
\textbf{Definition}\\
\toprule
API &
Application Programming Interface\\
\hline
C\&C &
Command-and-Control Server \\
\hline
CBC &
Classical Binary Classifier \\
\hline
COBYLA &
Constrained Optimization by Linear Approximation \\
\hline
DGA &
Domain Generation Algorithm\\
\hline
(D)DoS &
(Distributed) Denial of Service \\
\hline
DL &
Deep Learning \\
\hline
EstimatorQNN &
Estimator Circuit of a Quantum Neural Network \\
\hline
GCP & 
Google Cloud Platform \\
\hline
HQBC &
Hybrid Quantum Binary Classifier \\
\hline
HQDL &
Hybrid Quantum Deep Learning \\
\hline
HTC &
Hoeffding Tree Classifier \\
\hline
ML &
Machine Learning \\
\hline
MLP &
Multilayer Perceptron \\
\hline
NISQ &
Noisy Intermediate-Scale Quantum \\
\hline
P2P &
Peer-to-peer \\
\hline
Pegasos &
Primal Estimated sub-Gradient Solver for Support Vector Machines \\
\hline
SPSA &
Simultaneous Perturbation Stochastic Approximation \\
\hline
QSIEM & 
Quantum-enhanced Security Information and Event Management \\
\hline
QAOA & 
Quantum Approximate Optimization Algorithm \\
\hline
QDL &
Quantum Deep Learning\\
\hline
QESG &
Quantum Estimated Sub-Gradient \\
\hline
QGAN &
Quantum Generative Adversarial Networks \\
\hline
QML &
Quantum Machine Learning \\
\hline
QNN(C) &
Quantum Neural Network (Classifier) \\
\hline
QCA &
Quantum Cybersecurity Analytics \\
\hline
QSVC &
Quantum Support Vector Classifier\\
\hline
QSVM &
Quantum Support Vector Machine\\
\end{tabular}
\end{table}

\begin{table}[H]
\centering
\setlength{\tabcolsep}{1.2em}\setlength\extrarowheight{3pt}
\begin{tabular}{l|l}
\textbf{Term} &
\textbf{Definition} \\
\toprule
SamplerQNN &
Sampler Circuit of a Quantum Neural Network \\
\hline
SIEM &
Security Information and Event Management \\
\hline
SLSQP &
Sequential Least Squares Programming optimizer \\
\hline
SMO &
Sequential Minimal Optimization\\
\hline
SOAR &
Security Orchestration, Automation and Response \\
\hline
SPSA & 
Simultaneous Perturbation Stochastic Approximation\\
\hline
SGD &
Stochastic Gradient Descent \\
\hline
TLS & 
Transport Layer Security \\
\hline
VQC &
Variational Quantum Classifier\\
\end{tabular}
\end{table}

\newpage
\vspace{1em}
\bibliographystyle{unsrt}
\bibliography{main}

\newpage
\listoffigures
\newpage
\listoftables
\newpage
\listofreferencedlists

\end{document}